\documentclass[twocolumn,twoside,slac]{revtex4}
\usepackage{graphicx}
\usepackage{fancyhdr}
\begin{document}

\title{The CDF Data Handling System }

\author{Dmitry O. Litvintsev\footnote{for the CDF Data Handling group}}
\affiliation{FNAL, CD/CDF, Batavia, IL 60510, USA} 

\begin{abstract}

The Collider Detector at Fermilab (CDF) records
proton-antiproton collisions at center of mass energy of 2.0 TeV at
the Tevatron collider. A new  collider run, Run II, of the Tevatron
started in April 2001. Increased  luminosity will result in about 1~PB of data recorded on tapes in the next two years. Currently
the CDF experiment has about 260 TB of data stored on tapes.  This
amount includes raw and reconstructed data and their derivatives.
	The data storage and retrieval are managed by the CDF Data
Handling (DH) system. This system has been designed to accommodate the
increased demands of the Run II environment and has proven robust and
reliable in providing reliable flow of data from the
detector to the end user. This paper gives  an overview of the CDF Run
II Data Handling system which has evolved  significantly over the
course of this year. An outline of the future direction   of the
system is given.

\end{abstract}

\maketitle

\thispagestyle{fancy}

\section{INTRODUCTION}

The Collider Detector at Fermilab (CDF) is a general purpose detector 
at the Fermilab Tevatron \cite{cdf-detector}. The Tevatron, the world largest 
$\rm{p}\bar{\rm{p}}$ 
collider with the c.m.s. energy of about 2~TeV, has undergone a major 
upgrade for  Run II that started April 2001. 
The CDF detector has been equipped with a new tracking system, a TOF system, 
a new plug calorimeter and luminosity counters. The muon system coverage has been 
extended. The CDF trigger and  DAQ systems have been upgraded to accommodate 
10$\times$ increase in luminosity. 

The experimental program at CDF includes
search for Higgs boson, precision measurements of electroweak parameters,
 study of t-quark properties, QCD at large ${\rm Q}^2$, heavy flavor physics and 
search for phenomena beyond Standard Model.

During the first years of Run II the CDF collaboration plans to record about 1~PB
of data. This volume is more than 20 times the volume of the previous 
data recording, Run I. The sheer volume of data and increased analysis 
activity due to collaboration growth and extended physics potential 
constitute serious challenges for a data handling system. All data is accessed 
multiple times during several years of active analysis. 
Direct access storage devices (DASD) are not affordable for 
this data volume. Instead the data is archived to sequential media. 
Any of the data can be retrieved onto DASD into available space on a modified 
least recently used (LRU) basis. Users access the data only from disk. 

The purpose of the DH system is to collect, organize, 
archive and then make available the data to user analysis job.

\section{THE DATA ACCESS ORGANIZATION}

The CDF Run II data handling strategy is essentially an evolution of the 
Run I approach of successive filtering of events of interest from huge 
primary datasets produced at CDF production farm into smaller sub-samples 
relevant for individual analyses. The process of filtering of events 
and information stored per each event continues until the samples used 
to produce final physics results are obtained. These final samples could be 
small enough to be held on disk. 

CDF has adopted a hierarchical data organization with the dataset at the highest 
level and runsection at the lowest level of the structure. The dataset is the 
collection of events passing pre-defined set of Level-3 paths (primary and secondary datasets) 
or other selection criteria relevant for particular physics analysis (tertiary or derived datasets). 
Level-3 path is defined as AND of Level-1, Level-2 and Level-3 triggers. 

\begin{figure}[bth]
\centering
\includegraphics[width=80mm]{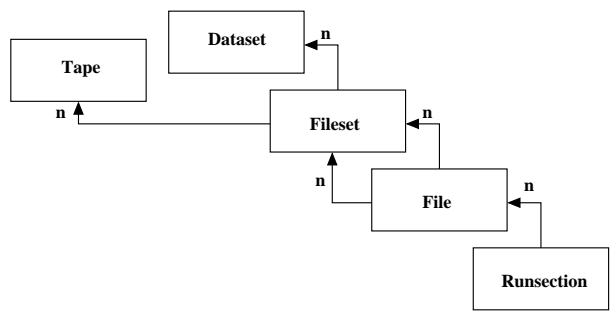}
\caption{Data access hierarchy}
\label{data-access}
\end{figure}

There are 50 pre-defined primary datasets at CDF. During the data taking events 
that belong to similar datasets are grouped into 8 streams. Grouping of datasets into the 
streams is done in such a way that event overlap between the streams is minimized 
and the fraction of the stream in any dataset is not small.

Runsections are the time intervals of data taking for which integrated luminosity  
is calculated. Typically, a runsection is defined every 30 seconds of 
data taking and contains on average 3,000 events. Events are written to 
files of about 1~GB in size. No runsection is split between two files. 
Groups of 10~GB or more worth of files form filesets to optimize tape I/O. 

At the beginning of the Run II the CDF was writing data to partitioned AIT-2 tapes, 
one fileset per partition. The individual data unit existing in the 
Data Handling system was therefore a fileset and not a file. 

Users access data by datasets following the datasets $\rightarrow$ filesets $\rightarrow$ files $\rightarrow$ runsections 
hierarchy. At the same time data handling access to data followed 
tape$\rightarrow$filesets$\rightarrow$files hierarchy. These two interlacing access patterns 
are shown in Figure~\ref{data-access}.

\subsection{Data Flow}

During data taking, the Consumer-Server/Logger (CSL) \cite{csl}
 receives events from the Level-3 PC farm at 20 MB/sec (75 Hz$\times$250 kB/event)
 and logs files to dual ported SCSI RAID array  disks at 20 MB/sec. 
These functions are performed on b0dau32, an SGI 2200 server dedicated to 
the CSL and located in CDF assembly building. A fileset-tape daemon 
running on another identical SGI 2200, fcdfsgi1, located 
in Feynman Computer Center, forms files into filesets and logs 
them to Mass Storage System (MSS) using the Enstore \cite{enstore} 
interface layer that provides access to network-attached 
tape drives in the STK robotic tape library.

Once the data are on tape and the calibrations are defined,
the raw data are fed to the CDF Production 
Farm \cite{farm} where they are reconstructed. 
After production, the data are split into the 
50 primary datasets. These datasets are written to tapes. 
The average Production Farm I/O throughput is about 30 MB/sec. 

\begin{figure}[t]
\centering
\includegraphics[width=85mm]{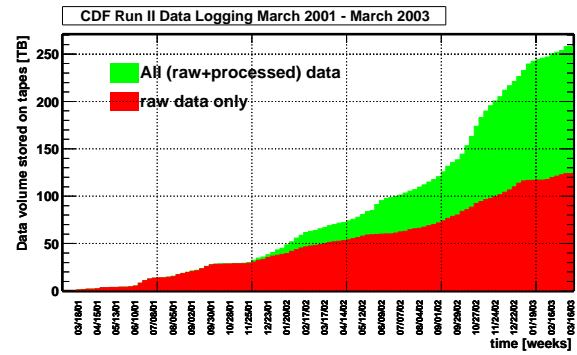}
\caption{Raw and produced data at CDF since the beginning of Run II} 
\label{data-size}
\end{figure}

The primary datasets are split into secondary datasets of interest for the 
physics analysis groups. Users create tertiary datasets or n-tuples 
using secondary datasets as inputs. Figure~\ref{data-size} 
shows the amount of raw and produced data logged to tape at CDF since 
the beginning of Run II. Some tapes containing older produced data 
were recycled to free up media.

\subsection{Data and Software Characteristics}

Some characteristics of the CDF data and analysis software:
\begin{itemize}
\item ROOT I/O as persistency mechanism
\item Typical raw event size is 250~kB
\item Typical produced event size is 350~kB, in  DST format (raw banks are kept)
\item PAD format or mini-DST format, ultimate event size 50-100~kB
\item N-tuple format 10-30 kB/event
\item Typical dataset size is $10^7$ events
\item Typical analysis job runs at 5 Hz on a 1~GHz 
PIII corresponding to few MB/sec input rate
\item Analysis jobs are CPU rather than network or I/O bound over Fast Ethernet
\end{itemize}

\section{CDF DATA HANDLING INFRASTRUCTURE}

\begin{figure}[t]
\centering
\includegraphics[width=85mm]{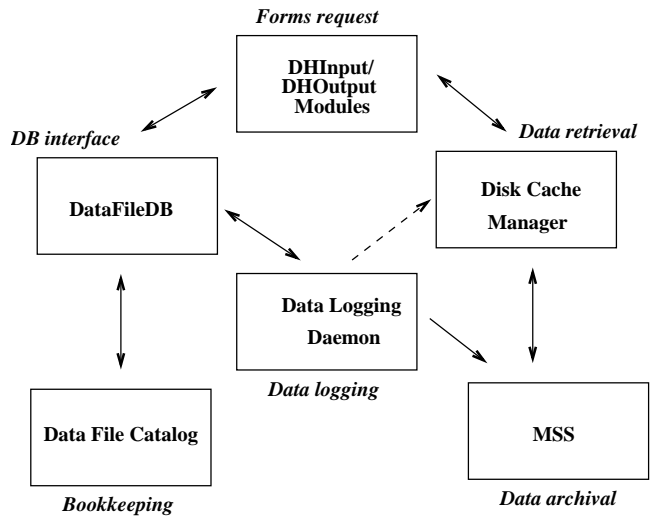}
\caption{The components of the CDF Data Handling system}
\label{dh}
\end{figure}

The DH system at CDF has several distinct components (Figure~\ref{dh})
each of which has a well-defined interface.

User specifies a request for data by dataset or other selection criteria based on 
meta-data information associated with the dataset, fileset or file via talk-to to a special 
DHInput module. The DHInput module translates the request into a list of filesets using information
available in the Data File Catalog via DB access layer provided by DataFileDB package.
The list of filesets is passed to a client API of the disk cache manager.
The client software contacts the disk cache manager server process to query the 
status of filesets in the list. The filesets available in the disk cache are processed while 
separate staging jobs are launched to copy missing filesets from MSS to disk. 

Raw and produced data are put to tape by data logging daemons that currently write directly 
into MSS using Enstore interface. 

\subsection{DH I/O modules}

An object oriented analysis framework, AC++ \cite{acpp} jointly developed by BaBar and CDF provides 
hooks to plug in high-level user interfaces to Data Handling system -- DHInput/DHOutput modules. 
The low-level API, the means to open, read and define internal structure of the data files is provided
by the Event Data Model \cite{edm}. The DHInput provides fast navigation through input data made possible 
by the direct access ROOT format of the CDF data. 

The DHOutput module writes out ANSI named files furnished with begin run record, empty runsections records, 
necessary for luminosity calculations, and makes entries in the Data File Catalog.

\subsection{Data File Catalog}

The Data File Catalog is a relational database that contains information about CDF datasets \cite{dfc}. 
It also contains physics-related bookkeeping information such as data 
quality, triggers and filters used, average and integrated luminosity, dynamic pre-scales values, etc.
The data access granularity of the DFC is a runsection, i.e. a group of events rather than individual 
events. The DFC table  structure reflects CDF data hierarchy depicted in Figure~\ref{data-access}.

The DFC tables store information about physical entities existing in the Data Handling system. 
Such entities are Dataset, Tape, Fileset, File and Runsection. Each entity has a corresponding primary 
table. The primary DFC tables are shown in Table I

\begin{table}[bth]\centering
\begin{tabular}{|l|l|} \hline
Entity     & Database table \\ \hline\hline
Dataset    & \verb+CDF2_DATASETS+ \\ \hline
Tape       & \verb+CDF2_TAPES+ \\ \hline
Fileset    & \verb+CDF2_FILESETS+ \\ \hline
File       & \verb+CDF2_FILES+ \\ \hline
Runsection & \verb+CDF2_RUNSECTIONS+ \\ \hline
\end{tabular}
\label{tab:primary}
\caption{Primary Data File Catalog tables}
\end{table}

Besides these primary tables there are secondary tables. These tables either keep certain relation 
between units, like for example parentage of the datasets or they keep history of changes, so called 
history tables, or they keep auxiliary entities like ranges or statuses or descriptions. 
There are 10 such tables (See Table II)

\begin{table}[bth]\centering
\begin{tabular}{|l|l|} \hline
Purpose                            & table name \\ \hline\hline
used to book dataset               & \verb+CDF2_DATASET_REGISTRIES+ \\ \hline
dataset parentage      & \verb+CDF2_PARENT_DATASETS+ \\ \hline
status table                                                & \verb+CDF2_DATASET_STATUSES+ \\ \hline
production version                         & \verb+CDF2_PROD_VERSION_DESCS+ \\
used to create dataset                                      & \\ \hline
 runsection ranges   & \verb+CDF2_RUNSECTION_RANGES+ \\
in a file                         & \\ \hline
trigger prescales                 & \verb+CDF2_FILE_LIVETIMES+ \\ \hline
dynamic trigger prescales         & \verb+CDF2_RUNSECTION_LIVETIMES+ \\ \hline
luminosity history                & \verb+CDF2_RUNSECTION_LUMINOSITIES+ \\ \hline
data quality bit description      & \verb+CDF2_DATA_QUALITY_DESCS+ \\ \hline
tape pool                         & \verb+CDF2_TAPEPOOLS+ \\ \hline
\end{tabular}
\label{tab:secondary}
\caption{Secondary Data File Catalog tables}
\end{table}

 Data File Catalog defines the following primary relationships among the 
entities:
\begin{itemize}
	\item a dataset has zero or more parent datasets
	\item a dataset contains zero or more filesets
	\item a dataset is contained on one or more tapes 
	\item tape pool may have zero or more tapes
	\item a tape has one parent tape pool
	\item a fileset has one parent dataset
	\item a fileset contains one or more files
	\item a file has one parent fileset
	\item a file has one parent dataset
	\item a file contains one or more runsection range
	\item a file contains zero or more average prescales
	\item a runsection contains zero or more dynamic prescales (livetimes)
	
\end{itemize}

There exists an Oracle implementation of the DFC tables used at Fermilab to keep track of centrally produced as 
well as secondary user data. There is also mSQL implementation of the DFC that can be set up and run 
at remote institutions if for some reasons either use of DFC located at Fermilab via network 
or installation and maintenance of Oracle DFC replica are not possible. Oracle and mSQL implementations 
are identical with the exception of latter not having integrity constraints and database triggers. 
Integrity constraints and the functionality of triggers are embedded in the DataFileDB library, 
a C++ DB access API. Recently the support for mySQL implementation of the DFC has been added. 

\subsection{DataFileDB API}

The information stored in the DFC is presented to the data processing algorithms as transient 
objects which are retrieved using compound keys. The management of mapping of persistent data to 
transient objects is provided by the common database interface manager layer\cite{chep2000_jbk}. 
This layer exists between the algorithm code and the code which reads directly from database 
tables. At the persistent storage end, it allows multiple back-end mapping classes to be plugged in 
and identified as data sources by character string at the run time. At the user end, it provides a
put/get/update/delete interface on top of a transient class for storage/retrieval/change/removal 
of objects of this class using a key. 

The mapping object creates a transient object from the data stored in the DFC using 
the key. Objects are cached by keys to prevent multiple database accesses from 
different algorithms for the same data. 

The DBManager\cite{chep2000_jbk} has two APIs. The back-end
persistent-to-transient mapping API,  {\it IOPackage}, is an abstract base class. 
In order to create new persistent-to-transient mapping class, or Mapper class, one 
has to derive this class from {\it IOPackage}. The user code sees the front-end API as template 
based. Transient classes are used as the template instantiation parameters of the 
\verb+Manager<object,key>+ class. The \verb+Manager<object,key>+ class has methods
such as put, get, update and remove to manipulate transient objects. 

The DataFileDB \cite{dfc} package is built on the top of DBManager  and 
provides all the code needed to manipulate the DFC from a C++ program.
It contains an implementation of the transient object classes, 
their associated keys and Mapper  classes. There are three  Mapper classes 
for each transient class corresponding to the three supported underlying relational database 
implementations -- Oracle (using OCI and OTL libraries), mSQL and mySQL. 
Information from any supported database implementation can be manipulated without code 
changes, the source database can be selected at the run time.

There are six classes representing rows of the DFC tables. Each of these classes, so called 
row-classes, has an interface that allows users to view the information held inside an object of the class.
Many of these objects contain the data collected from the secondary DFC tables like parent datasets 
or runsection ranges. Each transient object class has an associated key object class. 
The key object defines the {\tt WHERE} clause of the {\tt SQL} statement emitted 
by the Mapper back-end implementation for the corresponding transient object. 

Some of the row-objects have hierarchical views associated with them. 
The hierarchical views make calls down to  DBManager classes which 
perform connections to the database to do put, get, update or remove queries. 
E.g., the code to retrieve all the files belonging to the dataset, 
identified by dataset name identifier  {\tt aphysr} looks like this:

\begin{verbatim}
// make connection to database identified 
// by key "prd_dfc"
        DFCFileCatalogNode fc("dfc_prd");
// key class associated with file
        DFCFileKey key;
        key.setDatasetNameID("aphysr");
// typedef std::vector<DFCFile> DFCFiles;
// typedef Handle<DFCFiles> DFCFiles_var;
        DFCFiles_var files;	
        fc.findFiles(key,files);
\end{verbatim}

The \verb+findFiles+ method of hierarchical view \verb+DFCFileCatalogNode+ 
instantiates {\tt Manager} provided by {\tt DBManager} API 
with appropriate template parameters and arguments: 
\begin{verbatim}
        DFCFiles_mgr m("dfc_prd","DFCFiles");
        m.get(key,files);
\end{verbatim}

The front-end API is in {\verb+DFCFiles_mgr+} class which is \verb+typedef  Manager< DFCFiles,DFCFileKey >+.
The argument "dfc\_prd" identifies the entire
set of classes such as OTL or mSQL or mySQL Mappers to be used to perform data base operation.
 An ASCII text configuration 
file associates this string with the real database instance by including user, 
password, node name and class set name. 
The second argument instructs {\it IOPackage} factory which particular Mapper sub-class to instantiate. 
The object returned from the API is managed by a smart pointer ({\tt Handle<DFCFiles>}).

\begin{figure*}[t]
\includegraphics[width=145mm]{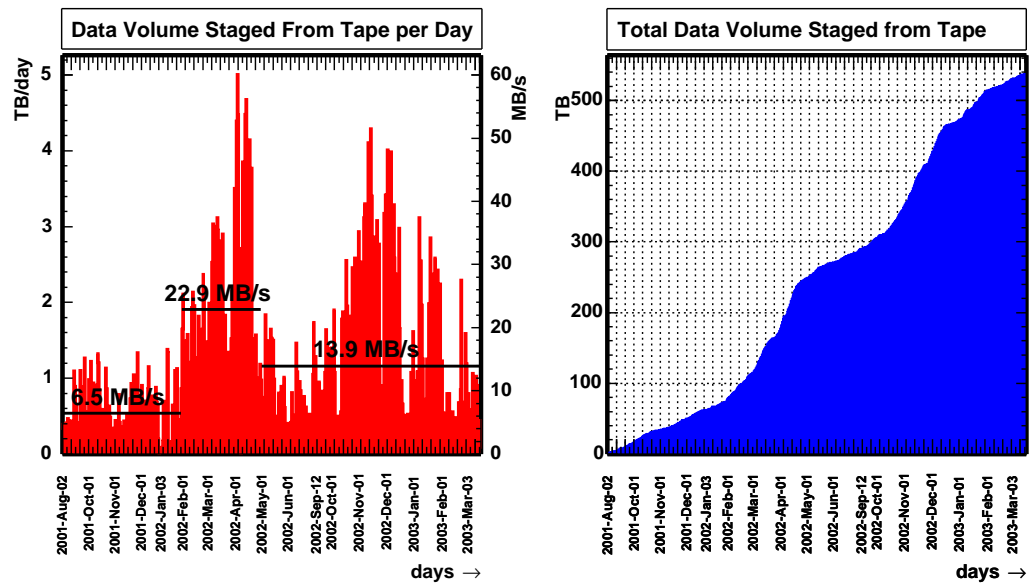}
\caption{Data volumes staged from tape on legacy DH system vs time}
\label{space}
\end{figure*}

\subsection{Disk Cache Management}

One of the central components of the CDF DH system is the Disk Inventory Manager (DIM) \cite{dim}. It acts as 
read/write cache in front of the MSS. The primary function of the DIM is to cache data from the tape library onto
a large collection of shared disks, secondary functions include automating the writing of new filesets 
onto tape, handling quotas for space and reservation management. 

The unit of space management is a fileset. DIM keeps track of fileset status and has no knowledge of datasets as such. This 
serves to decouple the manager from database system. The DIM is a client-server application with the communication 
between client and server via TCP/IP sockets. 

The server daemon is a multi-threaded program written to the POSIX C API using the worker pool model.
For better scalability it uses a dynamic threading system , starting more threads under load and eliminating them
when the load is reduced. The caching mechanism scores each fileset thet is not being reserved by users, 
just arrived or marked static based on reservations, time on disk and time since last usage. It deletes the fileset with the lowest score to make room for newly requested fileset, an LRU algorithm.

The initial design of the central analysis system was based on the idea of tight storage and CPU connection. The large computing machine, 
a SGI Origin 2000 with 128 300 Mhz MIPS R12000 CPUs, fcdfsgi2,  has been purchased and commissioned at the end 
of 1999. 
A 12 TB disk cache pool of Fiber Channel SCSI disk RAID arrays is managed by the DIM.

The DIM has originally been designed to work with attached disks on a large SMP. 
The SMP-centric model was first envisioned for Run II in 1998. In later revisions the model's 
inflexibility, cost/performance considerations  and single source upgrades 
were acknowledged and an extension towards incorporating commodity PCs and network 
distributed IDE RAID based disk storage was adopted 

While network nature of DIM allows it to handle network-attached storage, it has never been developed 
to fully support distributed caching. 

The CDF DH group evaluated an effort required to support DIM in distributed environment and decided in favor of adopting different 
cache layer -- dCache \cite{dcache}. 

dCache is a front-end disk cache for the large MSS. Originally conceived at DESY, dCache has been jointly 
developed by DESY and Fermilab Computing Division (CD)  for several years now. 
dCache uses network mounted disks to implement distributed 
data caches with user authentication. dCache provides file-based staging, making concept of fileset obsolete.

A client, requesting a file, contacts a dCache admin node and is authenticated. 
If the file is in any of the cache pools 
managed by the dCache admin server, the client is redirected to the pool 
containing the file. If the file is not on any of the 
pools is is staged from tape to pool with available disk space and client is redirected to that pool.

\subsection{Mass Storage System}

CDF has started Run II with a MSS based on an AML-2 robotic tape library with  cheap commodity AIT-2 tape drives,
SONY-CDX500C, directly attached to main CDF data logger and central analysis SMP, fcdfsgi2. Interface to 
MSS was written using CDF-specific software. The choice of tape technology turned out to be more difficult 
than anticipated. 

In May 2002, this tape system has been replaced with a Enstore Mass Storage System 
with dual STK Powderhorn 9310 robotic libraries equipped with 10 network-attached data center quality tape drives STK T9940A, 10~MB/sec
read/write rate each \cite{enstore}. CDF Enstore MSS is called CDFEN.

All existing data  were copied from AIT-2 
tapes to STK cartridges using the DH system running on the central analysis SMP. 

This year, STK T9940A are being replaced with STK T9940B tape drives. The total I/O rate of CDF tape system 
system becomes 10x30~MB/sec=300~MB/sec. This bandwidth is shared between 
raw data logging, Production Farms and a 600 CPU Central Analysis Farm (CAF) \cite{caf} for user analysis and 
legacy SMP system. Total data capacity is about 2 PB with 200 GB/tape cartridges.  The tape system I/O rate and 
volume capacity are sufficient for the Run IIa luminosity goals. 

Data delivery became stable and reliable. Figure~\ref{space} illustrates increased read rate from Enstore on legacy DH system.

\section{CURRENT DEVELOPMENT}

Having achieved operational stability and reliable data delivery at Fermilab, the CDF Data Handling 
group is looking forward to  further development of the Data Handling system towards 
providing data flow to world-wide distributed computing resources. 

The CDF has evaluated the D0 Data Handling system based on Sequential Access through 
Meta Data (SAM) system \cite{sam}. SAM is a complete Data Handling system that provides:
	\begin{itemize}
		\item Meta-data file catalog,
		\item Clustering the data onto tertiary storage in the manner
	               corresponding to access pattern
		\item Caching frequently accessed data on disk or tape,
		\item Organization of data request to minimize tape mounts
		\item A resource manager that estimates resources required for the 
	              file requests before they are submitted and, with this information,
		      makes administrative decisions concerning data delivery priorities
	\end{itemize}

The SAM infrastructure consists of a central data repository and a number of SAM-stations. A station 
consists of one or more computer nodes that share a data cache managed  by one or more station master 
nodes and accessible to consumer/producer nodes. Users submit jobs to computer in a stations specifying
the project their job will access. Using the file catalog meta-data, SAM translates the request into the 
list of files. If the requested files are available in the station cache they are sent to the requesting job. If some 
files are missing in the local cache, a station, following a set of rules will request the data from central repository or from neighboring stations. 

By design, the SAM is inherently a scalable, flexible Data Handling solution specifically 
tailored to accommodate distributed computing resources. This realization and 
the initial success of integrating SAM into CDF software \cite{chep2003_sam} resulted in creation of CDF SAM 
project in the framework of joint D0/CDF/CD project. 

The outcome of this project was a modified file catalog SAM schema that would allow 
to absorb the CDF Data File Catalog into SAM and interfacing of SAM data access layer with 
dCache for reading and writing data.

\section{CONCLUSION}

The CDF Collaboration has changed significantly its computing analysis system 
towards using globally distributed commodity CPU resources 
and network accessible IDE RAID arrays for disk caching. 

The choice of associated software and hardware components seems to be paying off remarkably well:
\begin{itemize}
	\item The {\it Enstore} generic interface to tape system that utilizes {\it data center quality} 
              drives allowed to achieve stable and robust operations in a very 
	      short period of time

	\item The {\it  network disk caching} management layer, dCache, runs successfully 
	    on commodity Linux file servers managing about 100 TB of distributed cache pools achieving
	    unprecedented TB/hour data delivery rate to analysis jobs running on CAF.

	\item An adaptation of the SAM as the first step towards GRID for CDF that 
	      ultimately allows off-site users to fully utilize their computing 
	      resources 
\end{itemize}

Most importantly CDF has a Data Handling strategy that allows scaling
with accumulated luminosity and increasing number CPUs and disks resources. 

The CDF Detector has achieved stable data taking. 
Stable detector operation complemented by a reliable Data Handling system 
will result in high quality and timely physics results. The first CDF Run II paper is in print, CDF is back in business. 

\begin{acknowledgments}

In conclusion the author wishes to thank members of CDF Data Handling group and 
 Fermilab Computing Division for their significant contribution to design,
implementation and operational support of the CDF Data Handling system.

\end{acknowledgments}


\end{document}